\documentclass{aa}
\usepackage{graphicx}
\usepackage{longtable}
\usepackage{txfonts}

\begin{document}
\renewcommand{\topfraction}{0.85}
\renewcommand{\bottomfraction}{0.7}
\renewcommand{\textfraction}{0.15}
\renewcommand{\floatpagefraction}{0.66}
\def\ecut{E_c=3.1(+0.5\,-0.4)_\mathrm{stat}\pm0.9_\mathrm{sys}}
\def\egamm{\Gamma=2.1\pm0.1_\mathrm{stat}\pm0.3_\mathrm{sys}}

\title{Observations of \object{Mkn~421} in 2004 with H.E.S.S. \\at large zenith angles}
\date{\today}

\author{F. Aharonian\inst{1}
 \and A.G.~Akhperjanian \inst{2}
 \and K.-M.~Aye \inst{3}
 \and A.R.~Bazer-Bachi \inst{4}
 \and M.~Beilicke \inst{5}
 \and W.~Benbow \inst{1}
 \and D.~Berge \inst{1}
 \and P.~Berghaus \inst{6} \thanks{Universit\'e Libre de 
 Bruxelles, Facult\'e des Sciences, Campus de la Plaine, CP230, Boulevard
 du Triomphe, 1050 Bruxelles, Belgium}
 \and K.~Bernl\"ohr \inst{1,7}
 \and C.~Boisson \inst{8}
 \and O.~Bolz \inst{1}
 \and I.~Braun \inst{1}
 \and F.~Breitling \inst{7}
 \and A.M.~Brown \inst{3}
 \and J.~Bussons Gordo \inst{9}
 \and P.M.~Chadwick \inst{3}
 \and L.-M.~Chounet \inst{10}
 \and R.~Cornils \inst{5}
 \and L.~Costamante \inst{1,20}
 \and B.~Degrange \inst{10}
 \and A.~Djannati-Ata\"i \inst{6}
 \and L.O'C.~Drury \inst{11}
 \and G.~Dubus \inst{10}
 \and D.~Emmanoulopoulos \inst{12}
 \and P.~Espigat \inst{6}
 \and F.~Feinstein \inst{9}
 \and P.~Fleury \inst{10}
 \and G.~Fontaine \inst{10}
 \and Y.~Fuchs \inst{13}
 \and S.~Funk \inst{1}
 \and Y.A.~Gallant \inst{9}
 \and B.~Giebels \inst{10}
 \and S.~Gillessen \inst{1}
 \and J.F.~Glicenstein \inst{14}
 \and P.~Goret \inst{14}
 \and C.~Hadjichristidis \inst{3}
 \and M.~Hauser \inst{12}
 \and G.~Heinzelmann \inst{5}
 \and G.~Henri \inst{13}
 \and G.~Hermann \inst{1}
 \and J.A.~Hinton \inst{1}
 \and W.~Hofmann \inst{1}
 \and M.~Holleran \inst{15}
 \and D.~Horns \inst{1}
 \and O.C.~de~Jager \inst{15}
 \and B.~Kh\'elifi \inst{1}
 \and Nu.~Komin \inst{7}
 \and A.~Konopelko \inst{1,7}
 \and I.J.~Latham \inst{3}
 \and R.~Le Gallou \inst{3}
 \and A.~Lemi\`ere \inst{6}
 \and M.~Lemoine \inst{10}
 \and N.~Leroy \inst{10}
 \and T.~Lohse \inst{7}
 \and A.~Marcowith \inst{4}
 \and C.~Masterson \inst{1,20}
 \and T.J.L.~McComb \inst{3}
 \and M.~de~Naurois \inst{16}
 \and S.J.~Nolan \inst{3}
 \and A.~Noutsos \inst{3}
 \and K.J.~Orford \inst{3}
 \and J.L.~Osborne \inst{3}
 \and M.~Ouchrif \inst{16,20}
 \and M.~Panter \inst{1}
 \and G.~Pelletier \inst{13}
 \and S.~Pita \inst{6}
 \and G.~P\"uhlhofer \inst{1,12}
 \and M.~Punch \inst{6}
 \and B.C.~Raubenheimer \inst{15}
 \and M.~Raue \inst{5}
 \and J.~Raux \inst{16}
 \and S.M.~Rayner \inst{3}
 \and I.~Redondo \inst{10,20}\thanks{now at Department of Physics and
Astronomy, Univ. of Sheffield, The Hicks Building,
Hounsfield Road, Sheffield S3 7RH, U.K.}
 \and A.~Reimer \inst{17}
 \and O.~Reimer \inst{17}
 \and J.~Ripken \inst{5}
 \and L.~Rob \inst{18}
 \and L.~Rolland \inst{16}
 \and G.~Rowell \inst{1}
 \and V.~Sahakian \inst{2}
 \and L.~Saug\'e \inst{13}
 \and S.~Schlenker \inst{7}
 \and R.~Schlickeiser \inst{17}
 \and C.~Schuster \inst{17}
 \and U.~Schwanke \inst{7}
 \and M.~Siewert \inst{17}
 \and H.~Sol \inst{8}
 \and R.~Steenkamp \inst{19}
 \and C.~Stegmann \inst{7}
 \and J.-P.~Tavernet \inst{16}
 \and R.~Terrier \inst{6}
 \and C.G.~Th\'eoret \inst{6}
 \and M.~Tluczykont \inst{10,20}
 \and G.~Vasileiadis \inst{9}
 \and C.~Venter \inst{15}
 \and P.~Vincent \inst{16}
 \and H.J.~V\"olk \inst{1}
 \and S.J.~Wagner \inst{12}}

\offprints{\\D. Horns, \email{Dieter.Horns@mpi-hd.mpg.de}}
\institute{
Max-Planck-Institut f\"ur Kernphysik, Germany%, P.O. Box 103980, D 69029 Heidelberg, Germany
\and
 Yerevan Physics Institute, Armenia%, 2 Alikhanian Brothers St., 375036 Yerevan, Armenia
\and
University of Durham, Department of Physics, U.K.%, South Road, Durham DH1 3LE, U.K.
\and
Centre d'Etude Spatiale des Rayonnements, CNRS/UPS, Toulouse, France% 9 av. du Colonel Roche, BP 4346, F-31029 Toulouse Cedex 4, France
\and
Universit\"at Hamburg, Institut f\"ur Experimentalphysik, Germany%, Luruper Chaussee 149, D 22761 Hamburg, Germany
\and
Physique Corpusculaire et Cosmologie, IN2P3/CNRS, Coll{\`e}ge de France, Paris, France%, 11 Place
%Marcelin Berthelot, F-75231 Paris Cedex 05, France
%APC, 11 Place Marcelin Berthelot, F-75231 Paris Cedex 05, France 
\thanks{UMR 7164 (CNRS, Universit\'e Paris VII, CEA, Observatoire de Paris)}
\and
Institut f\"ur Physik, Humboldt-Universit\"at zu Berlin, Germany%, Newtonstr. 15, D 12489 Berlin, Germany
\and
LUTH, UMR 8102 du CNRS, Observatoire de Paris, Section de Meudon, France % F-92195 Meudon Cedex, France
\and
Groupe d'Astroparticules de Montpellier, IN2P3/CNRS, Universit\'e Montpellier II, France%, CC85, Place Eug\`ene Bataillon, F-34095 Montpellier Cedex 5, France 
\and
Laboratoire Leprince-Ringuet, IN2P3/CNRS, Ecole Polytechnique, Palaiseau, France%, F-91128 Palaiseau, France
\and
Dublin Institute for Advanced Studies, Ireland%, 5 Merrion Square, Dublin 2, Ireland
\and
Landessternwarte Heidelberg, K\"onigstuhl, Heidelberg, Germany
\and
Laboratoire d'Astrophysique de Grenoble, INSU/CNRS, Universit\'e Joseph Fourier, France % BP 53, F-38041 Grenoble Cedex 9, France 
\and
Service d'Astrophysique, DAPNIA/DSM/CEA, CE Saclay, Gif-sur-Yvette, France
\and
Unit for Space Physics, North-West University, Potchefstroom , South Africa
\and
Laboratoire de Physique Nucl\'eaire et de Hautes Energies, IN2P3/CNRS, Universit\'es
Paris VI \& VII, France%, 4 Place Jussieu, F-75231 Paris Cedex 05, France
\and
Institut f\"ur Theoretische Physik, Lehrstuhl IV,
    Ruhr-Universit\"at Bochum, Germany%, D 44780 Bochum, Germany
\and
Institute of Particle and Nuclear Physics, Charles University, Prague, Czech Republic%,
%    V Holesovickach 2, 180 00 Prague 8, Czech Republic
\and
University of Namibia, Windhoek, Namibia
\and
European Associated Laboratory for Gamma-Ray Astronomy, jointly
supported by CNRS and MPG
}

\date{Received 8 April 2005/ Accepted 14 April 2005}
\abstract{
	\object{Mkn~421} was observed during a high flux state for nine nights in April
and May 2004 with the fully operational High Energy Stereoscopic System
(H.E.S.S.) in Namibia. The observations were carried out at zenith angles
of 60$^\circ$--65$^\circ$, which result in an average energy threshold of 1.5
TeV and a collection area reaching 2~km$^2$ at 10~TeV.  Roughly 7000 photons
from Mkn~421 were accumulated with an average gamma-ray rate of 8 photons/min.
The overall significance of the detection exceeds 100 standard deviations. The
light-curve of integrated fluxes above 2~TeV shows changes of the diurnal flux
up to a factor of 4.3. For nights of high flux, intra-night variability is
detected with a decay time of less than 1 hour.  The time averaged
energy spectrum is curved and is well described by a power-law with a photon
index $\egamm$ and an exponential cutoff at $\ecut$~TeV and an average integral
flux above 2~TeV of 3 Crab flux units.  Significant variations of the spectral
shape are detected with a spectral hardening as the flux increases.
Contemporaneous multi-wavelength observations at lower energies (X-rays and
gamma-rays above $\approx 300$~GeV) indicate smaller relative variability
amplitudes than seen above 2~TeV during high flux state observed in April
2004.
} 

\maketitle
	\section{Introduction} 
	Mkn~421 is a \lq BL Lac\rq\, type active galactic nucleus. The
broad-band spectral energy distribution is dominated by non-thermal emission
that is believed to be produced in a relativistic jet pointing towards the
observer. The high energy emission of this object has been studied by previous
observations carried out by northern hemisphere ground based Cherenkov
telescopes (Punch et al. \cite{punch}, Aharonian et al. \cite{2002A&A...393...89A}, Krennrich et al. \cite{2002ApJ...575L...9K}).
Observations of Mkn~421 from the southern hemisphere at large zenith angles benefit from 
considerable increase of the collection area at higher energies, which
results in a better temporal resolution at high energies and a
better sampling of the high energy part of the energy spectrum (Okumura et al. \cite{2002ApJ...579L...9O}). 

Besides the general interest of understanding the physics of the highly
relativistic plasma and its interaction with the ambient medium, the proximity
of Mkn~421 (z=0.031) makes it an interesting target to observe the effect of
pair-production of gamma-rays with soft (thermal) background photons as part of
the extragalactic background light. As the collection area for large zenith
angle observations exceeds two square kilometers at energies beyond 10~TeV, the observable energy
spectrum could eventually be extended beyond 20~TeV, which is important to probe the
mid-to-far infrared part of the extragalactic background light.

	\section{Observations and Data analyses} 
%1
	 H.E.S.S. (Hofmann et al. \cite{hofmann}) is an imaging
atmospheric Cherenkov detector dedicated to the ground based observation of
gamma-rays at energies above 100~GeV.  Situated in Namibia (23$^\circ$16'S
16$^\circ$30'E), the full array of four telescopes is operational since
December 2003. Each telescope has a mirror area of 107~m$^2$ and is equipped
with a camera consisting of 960 photomultiplier tubes (Vincent et al. \cite{vincent}).
The system has a field of view of 5$^\circ$ and allows to reconstruct the
direction of individual showers with a precision of better than 0.1$^\circ$.

%2
	 The H.E.S.S. observations reported here were carried out for typically
1--2~hours per night from MJD 53107.8 to MJD 53114.9 (April 12--19, 2004)
triggered by an increased level of X-ray emission detected by the
All-Sky-Monitor (ASM) onboard the RXTE satellite and increased
activity detected by the Whipple Cherenkov telescope (Krawczynski private
communication). Online analysis of the
H.E.S.S.  data revealed that the source was also active at TeV energies
motivating an extension of the observational campaign for roughly one week. In
the beginning of May (MJD 53134.8 corresponding to  May 8, 2004), the H.E.S.S.
array participated in a multi-wavelength campaign with overlap with
pointed X-ray observations with the RXTE satellite. 

%3
	The observation mode was employing all four telescopes, pointing with
0.5$^\circ$ offset in declination with sign alternating from run to run (28
minutes each run). The runs were selected according to general quality checks
such as absolute value of the trigger rate, relative changes in the rate and
performance of the cameras (number of operational pixels, homogeneity of the
acceptance). In total, 14.7 hours of good quality data were selected for the
analyses.  The zenith angle of the observations ranges from
60.3$^\circ$ at culmination to 65.4$^\circ$ with an average energy threshold at
1.5~TeV. 
%ADDED ACCORDING TO REFEREE'S REPORT
The average size of images and amplitude detected in individual
pixels at the given zenith angles is well below the saturation limit of
the H.E.S.S. cameras and read out electronics.

%4
	The data were calibrated following the general procedure (Aharonian et
al. \cite{2004APh....22..109A}) and standard data reduction (image cleaning, event reconstruction)
was applied (Aharonian et al. \cite{2004astro-ph0411582}).  The cuts used for background rejection
were identical to the ones used in previous analyses (e.g. Aharonian et al.
\cite{2004A&A...425L..13A}, \cite{2004astro-ph0411582}) with the exception of a loose angular cut
$\theta^2<0.05(^\circ)^2$ (compared to $\theta^2<0.02(^\circ)^2$ at smaller
zenith angles), where $\theta^2$ is the squared angular distance 
between source and reconstructed shower direction.  The relaxed angular 
cut compensates for the degrading angular resolution at
low elevations.  Its value was not optimised to result in a maximum
significance but rather chosen to ensure a high ($>80\,\%$) and 
constant acceptance  for gamma-rays.
The relative energy resolution at the observed zenith angle range 
expected from simulation is roughly 25~\%.

%5
For estimating the background, five \textit{off} regions identical 
to the \textit{on} region in size and
separation from the camera center were chosen (Aharonian et al. \cite{ahar01}). 
The overall signal detected from
Mkn~421 amounts to $N_\mathrm{on}=8978$ and $\langle N_\mathrm{off}\rangle
=1357.6$ averaged over the five background regions with a significance of
$S=114~\sigma$ using the likelihood ratio method derived in Li\,\&\,Ma (\cite{1983ApJ...272..317L}).

%6
Finally, in order to reconstruct the energy spectra, collection areas were 
calculated from simulated air showers taking the range of observed zenith
angles and the offset of the source to the camera centre into account.
	 The collection area derived for individual events (using linear
interpolation between logarithmic energy bins  and linear interpolation between
zenith angles) were used as weights to calculate the flux in bins of energy by
summing over the events in the \textit{on} and \textit{off} regions.
The effect of spill-over between adjacent energy bins due to the
instrument's energy resolution was compensated by using collection areas as a
function of reconstructed energy (Mohanty et al. \cite{1998APh.....9...15M}, Aharonian et al. \cite{2002A&A...393...89A}). 

Estimates of systematic uncertainties were derived by changing the atmospheric
transparency in the simulations and re-applying the analyses to the data. The
small variations observed in the cosmic ray induced event rate  during the
observations (see also next section) are consistent with smaller variations of 
the atmospheric transparency than assumed in the evaluation of the systematic
uncertainties. 

%7
%	 The analysis method used here does not fully employ the potential of
%the fine pixelation of the H.E.S.S. cameras for large zenith angle data. More
%advanced techniques will be developed in order to improve 
%the angular resolution and background rejection. The main benefit
%will be for the highest energy part of the spectrum, where the signal obtained
%is limited by the background fluctuations.

        \section{Energy spectrum from the Crab nebula at large zenith angles}

%1
	In order to check the procedure for the reconstruction of energy
spectra at large zenith angles, a dedicated data-set taken on the Crab nebula
for 1.8~hours at zenith angles ranging from 57$^\circ$ to 67$^\circ$ was
analysed. 

%2
% moved to previous section

%3
The fit of a power-law ($dN/dE=N_0 (E/\mathrm{TeV})^{-\Gamma}$) to the resulting energy spectrum of the Crab nebula 
yields 
$N_0=(2.9\pm0.4_\mathrm{stat}\pm0.7_\mathrm{sys})\times 10^{-11}$
ph/(cm$^2$\,s\,TeV), $\Gamma=2.6\pm0.1_\mathrm{stat}\pm0.3_\mathrm{sys}$,
 $\chi^2=7(8\,\mathrm{d.o.f.})$.
 The result is in good agreement with the recently published values 
given by the HEGRA collaboration (Aharonian et al. \cite{ApJCrab}). This gives
confidence in the reliability of the instrument's response and reconstruction 
technique at large zenith angles.

	\begin{figure}
	\includegraphics[width=1.0\linewidth]{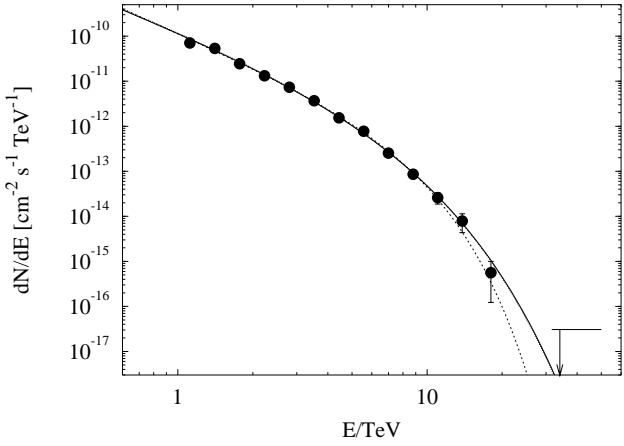}
	\caption{Differential energy spectrum of Mkn~421: The curvature 
is described by a power-law with a photon index of $\Gamma=2.1\pm0.1_\mathrm{stat}\pm0.3_\mathrm{sys}$
with an exponential cutoff at
$\ecut$~TeV (solid
line) or alternatively with a super-exponential cutoff ($\exp(-(E/E_c)^{1.6\pm0.3})$
at $E_c=(6.25\pm0.4_\mathrm{stat}\pm0.9_\mathrm{sys})$ TeV (dashed line). 
Between 30 and 50 TeV an upper limit with a 99~\% confidence level is
displayed. }
\end{figure}

\section{Energy spectra and light-curve from Mkn~421}

  The differential energy spectrum for Mkn~421 was calculated in the same
way as described above for the Crab nebula. The entire data-set 
was used to derive a differential energy spectrum.

	The final result for the energy spectrum of Mkn~421 is shown in Fig. 1.  As the
energy spectrum shows clear evidence for curvature, the collection areas were
calculated iteratively until the parameters of the model fit function converged.
The differential energy spectrum is well sampled between 1 and 15
TeV.  Between 15 and 30 TeV, a signal (corrected for 
an expected spill-over of 3 events) of 2.2$\sigma$ with $N_\mathrm{on}=31$
and $\langle N_\mathrm{off}\rangle=96/5=19.2$  events is seen. The collection
area for this energy bin exceeds 2~km$^2$ after cuts.  Between 30 and 50~TeV
an upper limit with a confidence level of 99~\% using the method of
Feldman~\&~Cousins (\cite{fc}) was calculated. All energy bins are centred on
the expected average energy for a given bin assuming a spectrum following a
power-law with an exponential cutoff.

	The result of a fit to the spectrum assuming a power-law with an
exponential cutoff (see solid line in Fig.~1), results in
$N_0=(1.55\pm0.08_\mathrm{stat}\pm0.4_\mathrm{sys})\times 10^{-10}$
ph/(cm$^2$\,s\,TeV), $\Gamma=2.1\pm0.1_\mathrm{stat}\pm0.3_\mathrm{sys}$,
$\ecut$~TeV, and  $\chi^2=16(10\,\mathrm{d.o.f.})$. Other model
fits using more complicated functions which have been commonly used in the past
to describe a curved spectrum were tested: a parabola in $\log(dN/dE)$ vs.
$\log(E)$  gives  $\chi^2=12.6(9\,\mathrm{d.o.f.})$ and a
super-exponential cutoff $\chi^2=11.7(9\,\mathrm{d.o.f})$. The
super-exponential cutoff fit with $dN/dE\propto
E^{-\Gamma}\times\exp(-(E/E_c)^{\alpha})$ with $\alpha=1.6\pm0.3$ 
(systematic uncertainty negligible)  is included
in Fig.~1 as a dashed line. According to the F-test, the probability of a
chance improvement of the $\chi^2$ for the super-exponential cutoff is 6.4~\%
which is not sufficiently low to give a preference to this fit function (the
same holds true for the parabolic fit function).

The cutoff energy $\ecut$~TeV found in the observations described here is fully
consistent with previous observations in the years 2000 and 2001 by HEGRA of
$3.6(+0.4\,-0.3)_\mathrm{stat}(+0.9\,-0.8)_\mathrm{sys}$~TeV (Aharonian et al.
\cite{2002A&A...393...89A}) and VERITAS $(4.3\pm0.3)~$TeV (Krennrich et al. \cite{2002ApJ...575L...9K}).

	 The diurnal integral fluxes above 2~TeV were calculated by taking the
sum over the inverse collection area for the events with energies above the
chosen value.  With the threshold energy raised to 2~TeV, this gives an
estimate of the flux which is independent of the variation of the actual energy
threshold within a night, at the expense of loosing count statistics. As an
additional benefit, the dependence on changes of the spectral shape are
negligible as the collection area is not strongly changing with energy
for energies well above the threshold. The
resulting light-curve is shown in Fig.~2.

The light-curve exhibits night-by-night variability in the first observing week
and resumes a lower flux in the later observation.  The peak diurnal average
flux reaches a value of about 5 times the flux observed from the Crab nebula.
The corresponding gamma-ray rate is sufficiently high to  probe intra-night
variability of the high energy end of the spectrum with unprecedented accuracy.  

During the observations of MJD 53113.8--53113.9 (April 18) and 53114.8--53114.9
(April 19), significant variations of the flux within these nights are detected.
The hypothesis of a constant flux during these nights results in
$\chi^2=40.6(7~\mathrm{d.o.f.})$ and
$\chi^2=28(7~\mathrm{d.o.f.})$  respectively.  The intra-night
variations as seen during MJD 53113.8--53113.9 (April 18) using bins of 14~min
width are shown as an inlay in Fig.~2.  In order to exclude variations
of the detector's response or changes in the atmosphere to be
responsible for the observed variations, the post-cut (after
applying the image shape cuts) cosmic ray rate has been checked for variability which is smaller than 2~\% in
relative root mean square (RMS) during this night.
The observed significant intra-night variability in the light-curve suggests a
decay time of less than 1 hour.

%%%%%%%%%%%%%%%%%%%%%%%%%%
\begin{figure}
	\includegraphics[width=\linewidth]{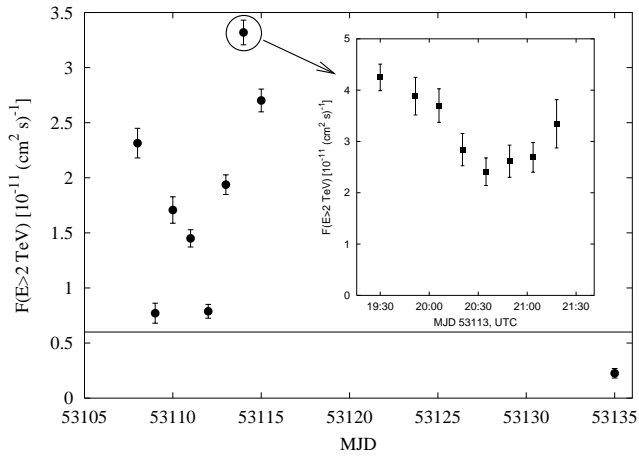} \caption{
The light-curve for Mkn~421 in April and May with diurnal integral 
flux values above 2~TeV. 
  For comparison, the horizontal line
indicates the flux observed from the Crab nebula. The inlay shows the
light-curve for the  night of MJD 53113 (marked with a circle and an arrow)
with the strongest indication  for intra-night variability in 14 min bins (0.01
d).} 
\end{figure}
%%%%%%%%%%%%%%%%%%%%%%%%%%

The light-curve is not corrected for possible
variations of the atmospheric transparency.  The atmospheric transmissivity
between the position of the air shower and the detector can be probed by the
rate of detected cosmic ray events, as the Cherenkov light traverses the same
aerosol layer which is believed to dominate possible temporal variations of the
atmosphere's transparency.  During the observations, the average cosmic ray
rates for individual runs varied between 113 and 143 Hz with an average of
132 Hz and RMS of 13 Hz. The post-cut rate shows even
less variation with 8~\% relative RMS. The observed variability of Mkn~421
is therefore clearly associated with the source and not a consequence
of temporal variations of the detector response or atmospheric transparency.
In principle, a correction of the
measured gamma-ray flux could be derived and applied to the light-curve. 
This was not done as the effect is small ($< 10\,\%$) in comparison
to the observed variability.

 In order to study variations of the spectral shape in various flux states,
diurnal spectra were calculated.  Given the correlation between the cutoff
energy $E_c$  and the photon index $\Gamma$ derived from the fit of a power-law
with an exponential cutoff (correlation coefficient $-0.955$ between $E_c^{-1}$
and photon index $\Gamma$), the power-law index was kept constant ($\Gamma=2$)
for the fits applied to spectra obtained for individual nights.  The fit was
applied to a fixed energy range from 1 to  10~TeV.  As a result, a hardening of
the spectrum or an increase of the cutoff energy is seen clearly in the
correlation of the cutoff energy $E_{c}$ and the integral flux above 2~TeV in
Fig.~3. A similar result was obtained when keeping the cutoff energy fixed and
letting the photon index vary freely.

\begin{figure}
\includegraphics[width=1.0\linewidth]{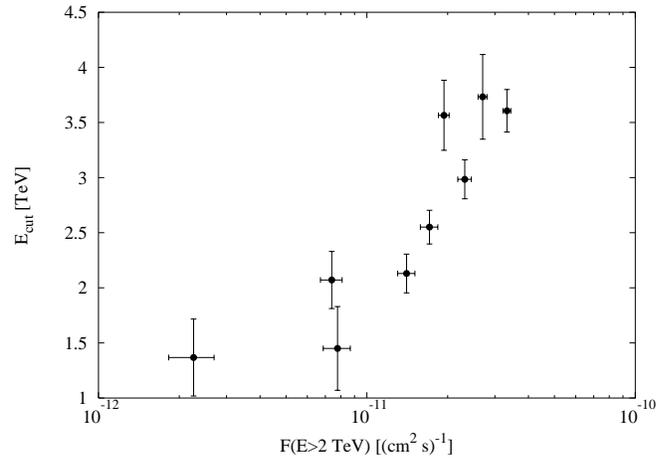}
\caption{The cutoff energy as determined from a fit with a fixed power-law
photon index $\Gamma=2$ versus integrated flux measured for $E>2~$TeV.
 Note, that the point with the lowest flux is from the May observations.} \end{figure}

	\section{Multiwavelength observations in April 2004}
 
%1
	During April 2004, a coordinated multi-wavelength campaign monitored
the activity of Mkn~421 in radio, optical, X-ray, and gamma-rays (Cui et al.
\cite{Cui}).  The source was seen to be active in X-rays where observations with the
array of proportional chamber units (PCU) onboard the RXTE satellite were
performed.  These observations were not simultaneous  with the observations
with the H.E.S.S.  array, but by combining the PCU with the ASM data a good
temporal coverage overlapping with the H.E.S.S. observations can be achieved.
An average of 4 counts/s were detected by the ASM during the first weeks of
April. This is sufficiently high to probe the activity of Mkn~421 during
individual ASM pointings (dwells). 

%2
 The Whipple 10 m Cherenkov telescope was observing Mkn~421 simultaneously 
with the RXTE pointings which were generally starting within a few hours after the H.E.S.S. observations took place. First preliminary results
of this campaign have been presented by Cui et al.~(\cite{Cui}). 

%3
  In Fig.~4, the different  observations are combined such that the observed
flux (or count-rate) is normalized to the average flux (or count-rate) during
the time between MJD 53107 and 53116.  The preliminary Whipple light-curve is
derived from the count-rate which is not corrected for different zenith angles
of observations. Generally, the Whipple observations were carried out at small
zenith angles (Cui et al. \cite{Cui}) and according to previous observations, 
the energy threshold is estimated to be $\approx 400$~GeV (Krennrich et
al. \cite{2002ApJ...575L...9K}).  The ASM light-curve and the $1~\sigma$ uncertainty band is obtained
by calculating the sliding average over five dwells. 

%The largest variability  amplitude is detected at energies above 2~TeV as
%seen with the H.E.S.S.  telescopes. The strongest flare observed on MJD 53113.9
%(April 18)
%is accompanied by a (37$\pm$20)\,\% increase of the X-ray emission as seen by
%the ASM. 

 Even though a detailed correlation analysis is difficult because of a large
fraction of the data not being simultaneous, it is interesting to note that the
strongest variability occurs at the highest energies, with a maximum relative
amplitude of the observed flux of $F_\mathrm{max}/F_\mathrm{min}=4.3\pm0.5$ as
compared to $1.7\pm0.2$ derived from the Whipple count rates and $2.05\pm0.02$
from the RXTE PCU observations.  As a measure of the variability the relative
RMS  of the measurement indicates a variability of $(17\pm6)$~\% seen by the
Whipple instrument and $(26\pm1)$~\% for the X-ray data taken with the 
PCU detectors, whereas for H.E.S.S.
the relative RMS value is ($51\pm8)$~\%.

\begin{figure}
 \includegraphics[width=\linewidth]{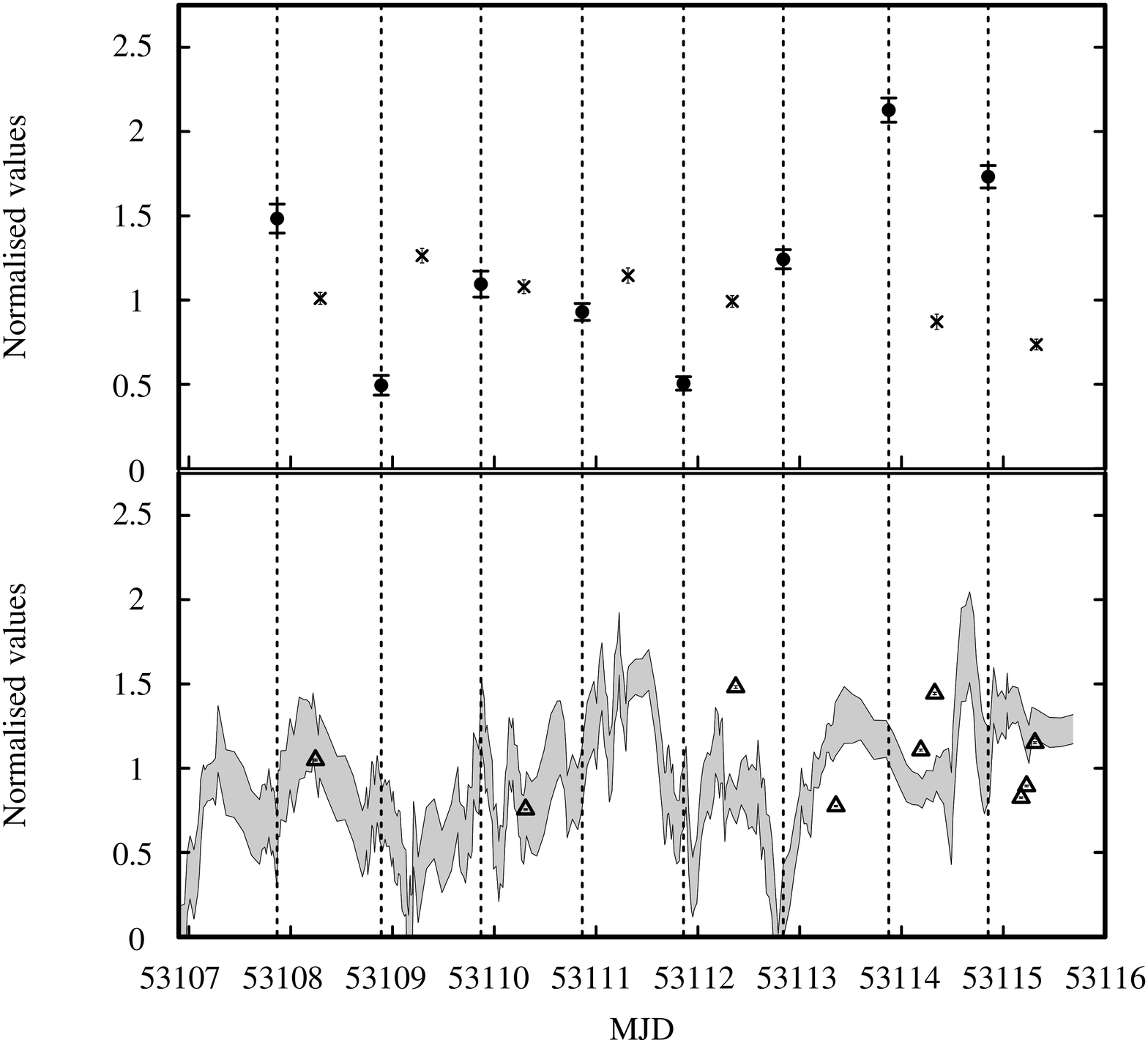}
\caption{
The relative changes  of the flux with respect to the average observed during
the multi-wavelength campaign in April 2004. Upper panel: H.E.S.S.  F(E$>$2
TeV) (filled circles), Whipple (crosses, preliminary count rate without
correction for varying zenith angles). Lower panel: PCU (open triangles,
count-rate), ASM (1$~\sigma$ error band, see text for details).  Whipple and
RXTE PCU data taken from Cui et al. (\cite{Cui}).}
 
\end{figure}

\section{Conclusion}
  The results on the time averaged energy spectrum of Mkn~421 presented here
confirm previous observations of the existence of a cutoff 
at $\ecut$~TeV  which is lower than the cutoff energy of 
$6.2\pm0.4_\mathrm{stat}(+2.9\,-1.5)_\mathrm{sys}$~TeV
observed from \object{Mkn~501} in 1997 (Aharonian et al. \cite{ahar99}). Given the similar red shift of these two objects, this would imply that the
cutoff in the Mkn~421 spectrum is 
intrinsic to the source and not
due to absorption.  With the observations carried out
at large zenith angles and increased collection area, a signal at the level of
2.2~$\sigma$ (corrected for the effect of spill-over) was detected 
between 15 and 30~TeV. This detection in
conjunction with the observation of a cutoff in the energy spectrum
intrinsic to the source is important for constraining the 
extragalactic background light at wavelengths beyond 10~$\mu$m. 
Furthermore,
the energy spectrum was observed to become harder as the integral
flux increases. It is not possible to discern whether the hardening is
a consequence of an increase of the cutoff energy from 1.5 to 3.5~TeV or 
a change in the power-law index as these two parameters are highly correlated.

The relative amplitude and variance of the variability observed above 2~TeV is 
significantly higher than observed at lower energies during the same
activity period by the Whipple telescope and the RXTE pointed X-ray detector. 
% ADDED ACCORDING TO REFEREE'S COMMENT
The sparse but similar sampling of the light curve among the different
observations (H.E.S.S., Whipple, and RXTE PCU) does not introduce a bias which
could explain the observed difference in the variability. 
%END

 The larger variability seen at multi-TeV energies compared to energies below
TeV is consistent with  the observed spectral changes, where the relative
increase of the flux at higher energies causes a hardening of the observed
spectra. 
%ADDED ACCORDING TO REFEREE'S 
Assuming the correlation between integral flux and cut-off energy 
indicated in Fig.~3 and using the integral flux above 2~TeV as measured
with H.E.S.S., the expected light curve above 400~GeV was calculated. 
As expected, the relative RMS scatter of the lower energy light curve is
roughly half of the value observed above 2~TeV well consistent with
the observed variability in the Whipple light curve.

\begin{acknowledgements} The support of the Namibian authorities
and of the University of Namibia in facilitating the construction and operation
of H.E.S.S. is gratefully acknowledged, as is the support by the German
Ministry for Education and Research (BMBF), the Max Planck Society, the French
Ministry for Research, the CNRS-IN2P3 and the Astroparticle Interdisciplinary
Programme of the CNRS, the U.K. Particle Physics and Astronomy Research Council
(PPARC), the IPNP of the Charles University, the South African Department of
Science and Technology and National Research Foundation, and by the University
of Namibia. We appreciate the excellent work of the technical support staff in
Berlin, Durham, Hamburg, Heidelberg, Palaiseau, Paris, Saclay, and in Namibia
in the construction and operation of the equipment.
\end{acknowledgements}

\end{document}